\documentclass[12pt]{article}
\usepackage{amssymb,amsthm,amsmath}
\usepackage{graphicx}
\usepackage{epstopdf}

\newcommand{\Eq}[1]{(\ref{#1})}
\newcommand{\Th}[1]{Th.~\ref{#1}}

\newcommand{\Sec}[1]{\S \ref{#1}}
\newcommand{\Fig}[1]{Fig.~\ref{#1}}

\newcommand{\InsertFig}[4]
{\begin{figure}[ht]
       \centerline{
         \includegraphics[width=#4]{./#1}
       }
       \caption{{\footnotesize  #2}
       \label{#3}}
\end{figure}}

\newcommand{\InsertFigTwo}[5] {
\begin{figure}[ht]
       \centerline{
			\includegraphics[width=#5]{./#1}
   		    	\hskip 0.25in
         	\includegraphics[width=#5]{./#2}
		}
       \caption{{\footnotesize  #3}
       \label{#4}}
\end{figure}}

\newcommand{\R}{{\mathbb{ R}}}

\newcommand{\eps}{\varepsilon}

\newtheorem{thm}{Theorem}

\newcommand{\sgn}{\mathop{\rm sgn}}

\title{Andronov-Hopf Bifurcations in Planar, Piecewise-Smooth,
Continuous Flows}
\author{
D.J.W.~Simpson and J.D.~Meiss\thanks
{
	We acknowledge support from the National Science Foundation
	through grant DMS-0202032.}\\
Department of Applied Mathematics\\
University of Colorado\\
Boulder, CO 80309-0526}

\date{\today}
\begin{document}
\maketitle

\begin{abstract}
\noindent
An equilibrium of a planar, piecewise-$C^1$, continuous system of differential equations that crosses a curve of discontinuity of the Jacobian of its vector field can undergo a number of discontinuous or border-crossing bifurcations. Here we prove that if the eigenvalues of the Jacobian limit to $\lambda_L \pm {\rm i} \omega_L$ on one side of the discontinuity and $-\lambda_R \pm {\rm i} \omega_R$ on the other, with $\lambda_L, \lambda_R >0$, and the quantity
$
	\Lambda = \lambda_L / \omega_L  -\lambda_R / \omega_R
$
is nonzero, then a periodic orbit is created or destroyed as the 
equilibrium crosses the discontinuity. This bifurcation is analogous
to the classical Andronov-Hopf bifurcation, and is supercritical if $\Lambda < 0$ and subcritical if $\Lambda >0$.

PACS: 02.30.Oz; 05.45.-a
\vspace*{1ex}
\noindent
\end{abstract}

\section{Introduction} \label{sec:Introduction}

Many dynamical systems can be modeled by a system of differential equations ${\dot z} = F(z)$ with vector field $F:M \rightarrow \R^n$ on an $n$-dimensional manifold $M$. Such a system is {\em piecewise-smooth continuous} (PWSC) if $F$ is everywhere continuous and is smooth except on the boundaries of countably many regions where it has a discontinuous Jacobian, $A = DF$. These boundaries are called the {\em switching manifolds}.

Piecewise-smooth dynamical systems are encountered in a wide variety of fields. Examples include vibro-impacting systems and systems with friction \cite{Br99, WiKr00}, switching circuits in power electronics \cite{BaVe01}, relay control systems \cite{ZhMo03} and physiological models \cite{KeSn98}.


Many bifurcations caused by discontinuities have been previously studied. Feigin
first studied period-doubling bifurcations in piecewise continuous systems and gave them the name ``$C$-bifurcations" for the Russian word {\em \v{s}vejnye} for ``sewing," and this term is often used more generally for any bifurcations that result from a discontinuity \cite{BeFe99, BeGa02, LeNi04, Lei06}. The bifurcations caused by the collision of a fixed point of a mapping with a discontinuity were studied by \cite{NuYo92} and given the name ``border-collision" bifurcations. Other examples of $C$-bifurcations include discontinuous saddle-node bifurcations (see \Sec{sec:db}), {\em grazing bifurcations}, where a periodic orbit touches a switching manifold, and {\em sliding bifurcations}, where a trajectory moves for some time along the switching manifold.

In this paper we study the bifurcation analogous to the Andronov-Hopf bifurcation that arises when an equilibrium encounters a switching manifold. Recall that in the classical Hopf bifurcation, a periodic orbit is generically created when an equilibrium has a pair of complex eigenvalues cross the imaginary axis \cite{Ku04,MaMc76}.
We will obtain a similar result for the case that an equilibrium of a planar, piecewise-$C^1$  continuous system crosses a switching manifold in such a way that its eigenvalues ``jump" across the imaginary axis. This theorem requires several nondegeneracy conditions; in particular, the criticality of the bifurcation is determined by the linearization of the critical equilibrium. This is to be contrasted with the classical smooth theorem where the distinction between supercritical and subcritical bifurcations depends upon cubic terms. Our result extends the result given in \cite{FrPo97} that applies to piecewise-linear systems. 

A related theorem was proved in \cite{ZoKu06} for the case of a piecewise-$C^2$ system (not necessarily continuous) for the case that the equilibrium is assumed to be fixed on the switching manifold, but its left and right eigenvalues change with a parameter.

\section{Discontinuous Bifurcations}\label{sec:db}

Piecewise-smooth, continuous odes may contain bifurcations that do not exist in
smooth systems. For instance, if an equilibrium crosses a switching 
manifold as a system parameter is continuously varied, we expect the eigenvalues to change discontinuously because of the discontinuity in the Jacobian. In this case the equilibrium may disappear or its stability may change, this is a {\em discontinuous} or {\em border-collision bifurcation} \cite{Lei06}.

Consider a PWSC system that depends upon a parameter $\mu$ and suppose that $z^*(\mu)$ is an equilibrium that lies on a switching manifold at $\mu =0$. Furthermore, suppose that the switching manifold is codimension-one and is smooth at the point $z^*(0)$. Then, without loss of generality, we can choose coordinates so that $z^*(0) = 0$ and the unit vector $\hat{e}_1$ is the normal vector to the switching manifold. To reflect these choices, let $z = (x,y)$, with $x$ representing the first component, and $y$ the remaining $n-1$ components.

As with smooth systems, knowledge of local behavior is gained by computing the Jacobian, $A(z) = DF(z)$, of the equilibrium near the bifurcation. Though the Jacobian is not defined at $z^*(0) = (0,0)$, the limits $A_L = \lim_{x \to 0_-} A(x,0)$ and $A_R = \lim_{x \to 0_+} A(x,0)$ do exist since $F$ is piecewise smooth. In this case, continuity of the system implies that all of the columns of $A_L$ and $A_R$ must be equal except for the first. The system is PWSC if the first columns do indeed differ.

Feigin gave one classification of border-collision bifurcations according to the spectra of $A_{L,R}$ \cite{BeGa02}. In particular, let $s_{L,R}$ be the number of negative, real eigenvalues of $A_{L,R}$, respectively. It is not hard to see that if $s_L + s_R$ is even the equilibrium generically persists and crosses the switching manifold, while if $s_L + s_R$ is odd the equilibrium does not cross the boundary but is destroyed in an analogue of the saddle-node bifurcation. Similar results are valid for maps where a simple criterion for a discontinuous period-doubling bifurcation can also be given \cite{BeFe99}.

Even when the stability of an equilibrium does not change upon collision with a switching manifold, its basin of attraction can undergo a dramatic change. In the piecewise linear case, orbits of a nominally stable equilibrium can be unbounded; this has been called a ``dangerous border-collision bifurcation" \cite{HaAb04}.

This paper is concerned with the planar case when $A_L$ and $A_R$
correspond to focus-type equilibria of opposing stability.
The discontinuous bifurcation arising in this situation displays
similar properties to an Andronov-Hopf bifurcation in a smooth system.

As an example, consider the piecewise-linear continuous system:
\begin{equation}\begin{split}\label{eq:example}
 \dot{x} &= -x - |x| + y \;,\\
 \dot{y} &= -3x + y - \mu \;.
\end{split}\end{equation}
For a given value of $\mu$, the system has a unique equilibrium,
namely a stable focus at $(-\mu,~-2\mu)$ when $\mu<0$
(with eigenvalues, $-\frac{1}{2} \pm \frac{\sqrt{3}}{2} i$)
and an unstable focus at $(-\frac{\mu}{3},~0)$ when $\mu>0$
(with eigenvalues, $\frac{1}{2} \pm \frac{\sqrt{11}}{2} i$).
The $y$-axis is a switching manifold and the equilibrium crosses this manifold at the origin when $\mu=0$ and changes stability.
Fig.~\ref{fig:example} shows phase portraits of \Eq{eq:example}
for negative and positive values of $\mu$.
A stable periodic orbit exists for all positive values
of $\mu$ and grows in size with $\mu$.
This situation is analogous to a supercritical Andronov-Hopf bifurcation
in a smooth system.

\InsertFigTwo{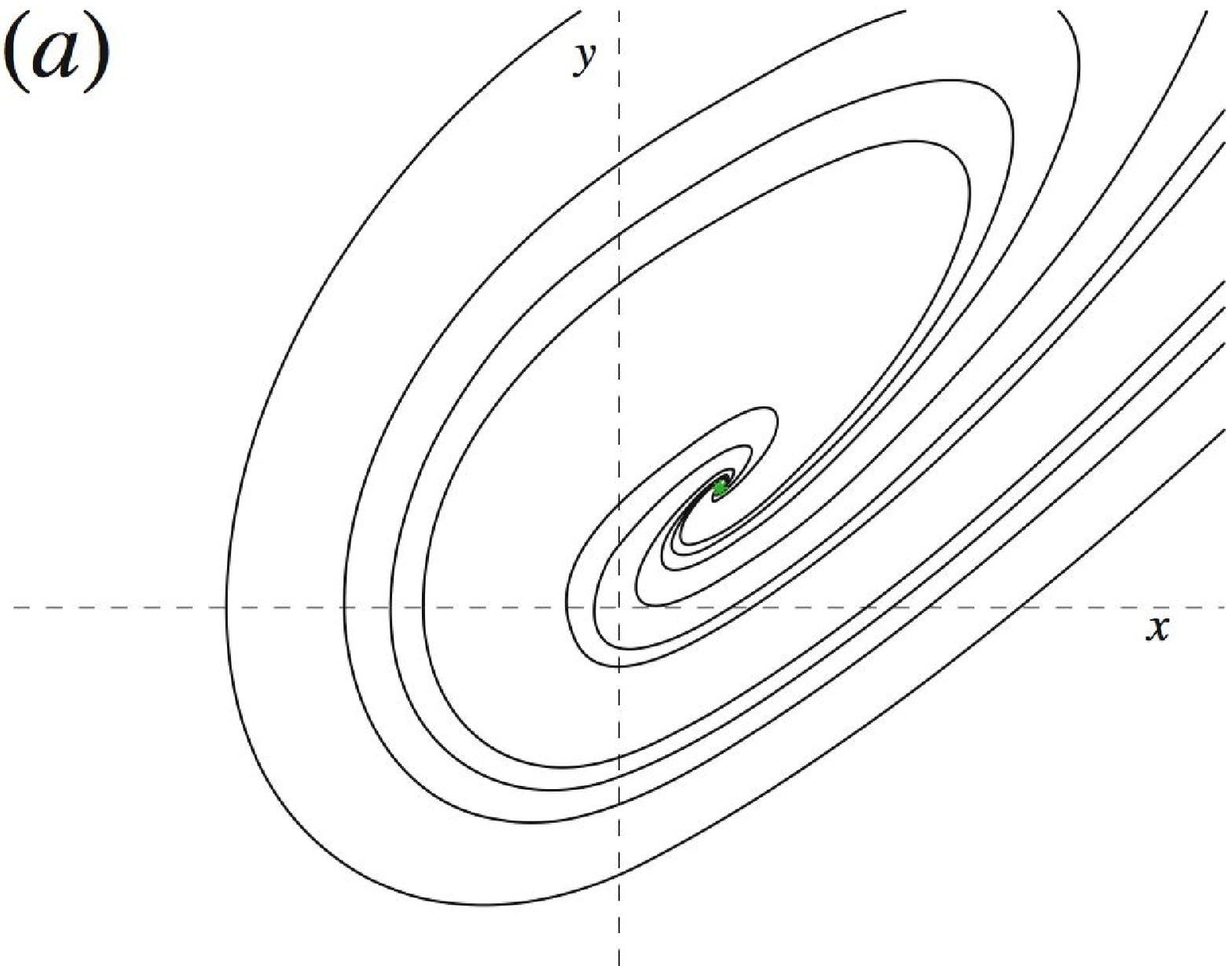}{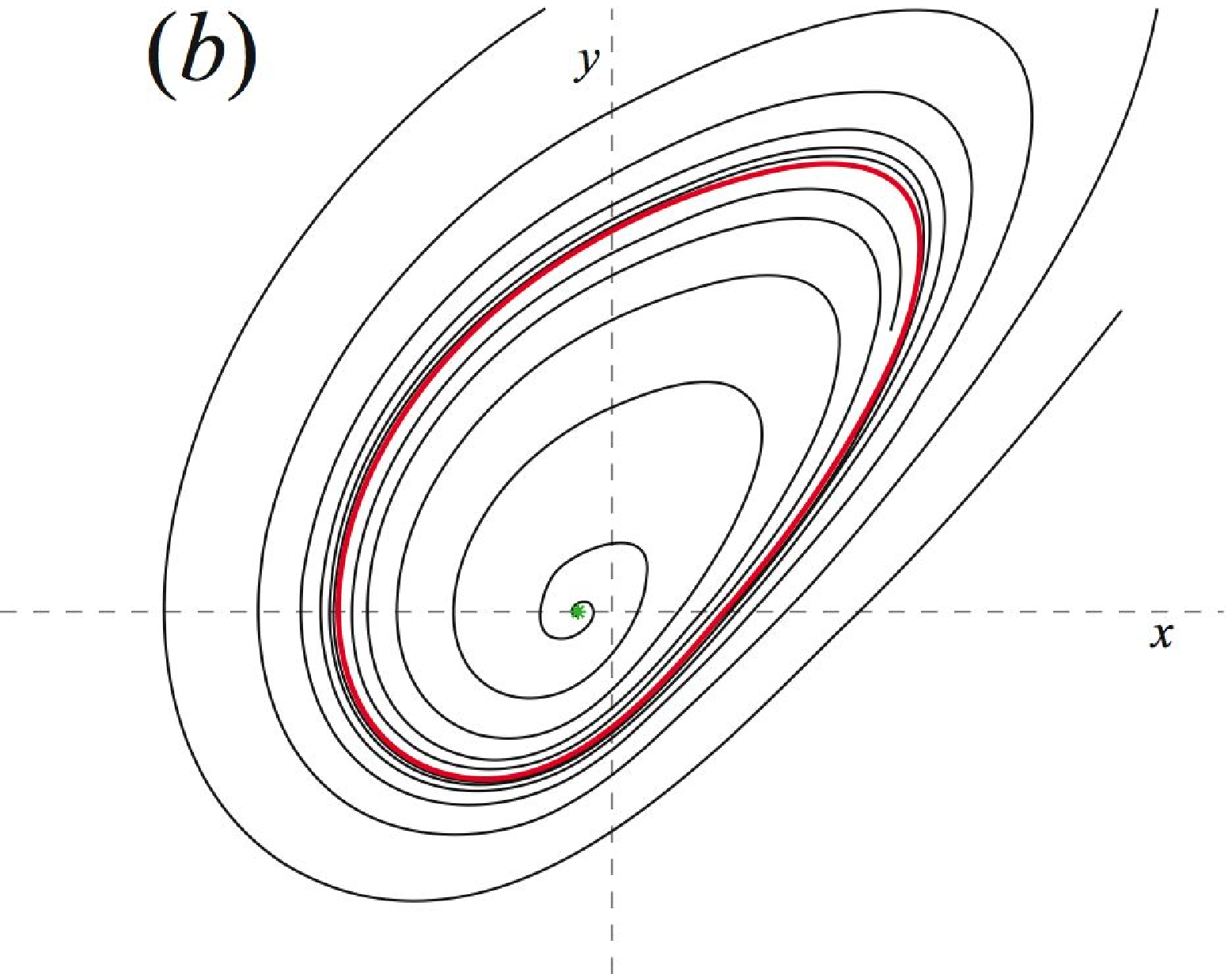}{Phase portraits of \Eq{eq:example}.
(a) $\mu=-0.01$. (b) $\mu=0.01$.}{fig:example}{6cm}

\section{Normal Form}

More generally, consider a two-dimensional, piecewise-$C^k$, continuous system
of ordinary differential equations in $z = (x,y) \in \R^2$, with $k \ge 1$.
Assume that there is an equilibrium,  $z^*(\mu)$, that crosses a switching manifold as a parameter $\mu$ is varied. If we assume this crossing occurs at a
differentiable point on a switching manifold, and that this manifold remains locally differentiable under small variations of $\mu$, we may further assume that the switching manifold coincides with, say, the $y$-axis and the crossing occurs at the origin when $\mu$ is, say, zero. Such a system can be written as:
\begin{equation}\label{eq:general}
	\dot z = \left\{ \begin{array}{ll} 
					F_L(x,y;\mu) \;, & x \le 0 \\
					F_R(x,y;\mu) \;, & x \ge 0 
					\end{array} \right. \;,
\end{equation}
where $F_L$ and $F_R$ are $C^k$ vector fields with $k \ge 1$. Since the vector field is continuous,
$F_L(0,y;\mu) \equiv F_R(0,y;\mu)$. We also assume that $F_L(0,0;0) = F_R(0,0;0) = 0$ to ensure that $z^*(0) = 0$. Let $j \in \{L,R\}$ represent the index for the two pieces of \Eq{eq:general}, so that we can rewrite the system as $\dot{z} = F_j(z;\mu)$ where $j = L$ when $x \le 0$ and $j= R$ when $ x \ge 0$. 

Expansion of each part of \Eq{eq:general} about the origin yields
\begin{equation}\label{eq:taylor}
\begin{split}
	\dot x = p(\mu)\mu + a_j(\mu)x + b(\mu)y + o(x,y) \;,\\
	\dot y = q(\mu)\mu + c_j(\mu)x + d(\mu)y + o(x,y) \;,
\end{split}
\end{equation}
where $p,q,a_j,b,c_j$, and $d$ are real-valued $C^{k-1}$ functions of $\mu$, for $j \in \{L,R\}$, and the correction terms go to zero faster than the first power of $x$ or $y$. The factor $\mu$ in each of the constant terms makes explicit the fact that the constant terms are zero when $\mu = 0$. The Jacobian matrices of the left and right systems in \Eq{eq:taylor} at the origin are
\begin{equation}\label{eq:matrices}
	A_j(\mu)= DF_j(0,0;\mu) =
					\left[ \begin{matrix} 
						a_j(\mu) & b(\mu) \\
						c_j(\mu) & d(\mu) 
					\end{matrix} \right] \;,
	\quad j \in \{L,R\} \;.
\end{equation}
We suppose $A_L(0)$ and $A_R(0)$ have eigenvalues
\begin{equation}\label{eq:eigen}
	\lambda_L \pm i\omega_L \;, \mbox{ and } -\lambda_R \pm i\omega_R \;,
\end{equation}
respectively, where
$\lambda_L,\lambda_R,\omega_L,\omega_R > 0$.
Notice this assumption implies that $c_j(0)$ and $b(0) \ne 0$ for
otherwise $A_j(0)$ would be triangular and have real
eigenvalues. Moreover note that $\det A_j(0) =\lambda_j^2 + \omega_j^2 > 0$. Finally, since continuity implies that $b(\mu)$ is the same for both matrices, the direction of
rotation for each subsystem must be the same.

The equilibrium of \Eq{eq:taylor} that lies at the origin when $\mu = 0$ will
non-tangentially cross
the switching manifold $y = 0$ if the parameter
\begin{equation}\label{eq:gamma}
	\Gamma \equiv p(0)d(0) - q(0)b(0) \ne 0 \;.
\end{equation}
In this case, by rescaling the parameter $\mu \mapsto \Gamma \mu$, we can effectively set
\begin{equation*}
	\Gamma = 1 \;.
\end{equation*}

Two other simple transformations can be done to simplify the coefficients
of \Eq{eq:taylor}. First, since $b(0) \neq 0$, we may use the transformation $y \mapsto b(0)y$, to set $b(0) = 1$. This implies that when $\mu$ is small, the rotation direction is clockwise.

Second, we may perform a $\mu$-dependent shift of $y$ to set $p(\mu) \equiv 0$ for small $\mu$.\footnote{ 
	This simplifies the proof of \Th{thm:dhb} in \Sec{sec:theorem}; in
	particular allowing us to center the Poincar\'e map \Eq{eq:poincare}, at the
	origin, instead of at some $\mu$-dependent point on the $y$-axis.
}
To see this, note that the implicit function theorem implies that 
there is a neighborhood of the origin in $(y, \mu)$ such that there is a unique $C^k$ function $\tilde{y}(\mu)$ that satisfies $\dot{x}(0,\tilde{y}(\mu);\mu)  \equiv 0$ with $\tilde{y}(0) = 0$. This follows because $\dot{x}(0,y;\mu)$ is $C^k$, $\dot{x}(0,0;0) = 0$ and $D_y \dot{x}(0,0;0) = b(0) = 1$. Thus the transformation $y \mapsto y - \tilde{y}(\mu)$, effectively sets $p(\mu) \equiv 0$ for all sufficiently small $\mu$. It is easy to see that the remaining coefficients $q,a_j,b,c_j$, and $d$ are still $C^{k-1}$ functions of $\mu$ (for small $\mu$) and are unchanged at $\mu = 0$.

The system \Eq{eq:taylor} now becomes
\begin{equation}\label{eq:dhb}
	\left[ \begin{array}{c} \dot{x} \\ \dot{y} \end{array} \right]
	=  \left[ \begin{array}{c} 0 \\ q(\mu) \end{array} \right] \mu 
	+ A_j(\mu) \left[ \begin{array}{c} x \\ y \end{array} \right]
	+ o(x,y) \;,\\
\end{equation}
for small enough $\mu$. Here we have $b(0)=1$ and since $p(0) =0$ we have $\Gamma = -q(0) = 1$ so that $q(0) = -1$. Note that the eigenvalues of $A_L(0)$ and $A_R(0)$ are still given by \Eq{eq:eigen}.

Let $x_L^*(\mu)$ be the $x$-component of the equilibrium when it exists in the left half-plane, and $x_R^*(\mu)$ in the right half-plane. It is easy to see that
\begin{equation}\label{eq:equil}
	x_j^*(\mu)  =  -\frac{\mu}{\lambda_j^2+\omega_j^2} + 
			o(\mu) \quad j \in \{L,R\}\\
\end{equation}
Thus when $\mu$ is small and positive [negative] the
equilibrium is located in the left [right] half-plane
and is a repelling [attracting] focus.

\section{Andronov-Hopf-like Bifurcations}\label{sec:theorem}

We will show that, as with an Andronov-Hopf bifurcation in a smooth system,
a periodic orbit of \Eq{eq:dhb} is created at $\mu = 0$ and grows in
amplitude as $\mu$ is either increased or decreased, given a single nondegeneracy condition.
Recall that the nondegeneracy condition for the smooth case is a cubic coefficient in the normal form \cite{Ku04, Wi03}. The sign of this coefficient also governs
the stability of the limit cycle, i.e., the criticality of the Hopf bifurcation.
For the discontinuous analogue that we are studying, the nondegeneracy condition depends only upon the eigenvalues of the linearized system.

\begin{thm}\label{thm:dhb}
Suppose that the vector field \Eq{eq:general} is continuous and piecewise $C^k$, $k \ge 1$, in $(x,y,\mu)$, and has an equilibrium that transversely crosses a one-dimensional switching manifold when $\mu = 0$ at a point $z^*$ where the manifold is $C^k$. Suppose further that as $\mu \to 0_+$ the eigenvalues of the equilibrium approach $\lambda_L \pm i \omega_L$ and as $\mu \to 0_-$ they approach $-\lambda_R \pm i\omega_R$, where $\lambda_L,\lambda_R,\omega_L,\omega_R > 0$.
Let
\begin{equation}\label{eq:Lambda}
 \Lambda \equiv \frac{\lambda_L}{\omega_L} - \frac{\lambda_R}{\omega_R} \;,
\end{equation}
denote the criticality parameter.

Then if $\Lambda < 0$ there exists an $\eps > 0$ such that
for all $0 < \mu < \eps $ there is an attracting periodic orbit whose radius is $O(\mu)$ away from $z^*$, and for $-\eps < \mu < 0$ there are no periodic orbits near $z^*$. 

If, on the other hand, $\Lambda > 0$, there exists an $\eps >0$ such that
for all $-\eps < \mu < 0 $ there is a repelling periodic orbit whose radius is $O(\mu)$ away from $z^*$, and for all  $0 < \mu < \eps$ there are no periodic orbits near $z^*$. 
\end{thm}

In order to prove \Th{thm:dhb} we will use the transformed system \Eq{eq:dhb} so that the switching manifold becomes (locally) the $y$-axis. The periodic orbit will be obtained as a fixed point of a Poincar\'e map, $P$, of the positive $y$-axis to itself.
This map is obtained as the composition of two maps $P_L$ and $P_R$ that
follow the flow in the left and right half-planes respectively.
We must prove that there is a neighborhood of the origin in which these two maps are well-defined. A fixed point of $P$ will be shown to exist using the implicit function theorem. This is essentially the same approach as that used in \cite{MaMc76} to prove the smooth Hopf bifurcation theorem and in \cite{FrPo97} to prove the same result for piecewise-linear systems.

It suffices to consider only $\mu \ge 0$, for when $\mu < 0$ the transformation
$(x,\mu,t) \mapsto -(x,\mu,t)$ produces a new system displaying the same properties
as those listed for \Eq{eq:dhb}. The signs of $\mu$
and $\Lambda$ become reversed and the stability of any periodic
orbits is flipped because we are reversing the direction of time.

\proof
\subsubsection*{Step 1: Define $P_L$ and $P_R$}

As previously argued, without loss of generality we may consider the $C^k$ system \Eq{eq:dhb}.
Consider the ``left" and ``right" systems (taken as though they were separately valid for both signs of $x$):
\begin{eqnarray}
	\left[ \begin{array}{c} \dot{x} \\ \dot{y} \end{array} \right]
	& = & \left[ \begin{array}{c} 0 \\ q(\mu) \end{array} \right] \mu
	+ A_L(\mu) \left[ \begin{array}{c} x \\ y \end{array} \right]
	+ o(x,y) \label{eq:dhbl} \\
	\left[ \begin{array}{c} \dot{x} \\ \dot{y} \end{array} \right]
	& = & \left[ \begin{array}{c} 0 \\ q(\mu) \end{array} \right] \mu
	+ A_R(\mu) \left[ \begin{array}{c} x \\ y \end{array} \right]
	+ o(x,y) \label{eq:dhbr}
\end{eqnarray}
Denote the flows of these equations by $\varphi_t^L(x,y;\mu)$ and $\varphi_t^R(x,y;\mu)$, respectively, for $(x,y,\mu)$ in some neighborhood of the origin.

For sufficiently small $\delta_j$ and $\delta_{\mu_j} > 0$, we will define 
the first return map
$$ 
	P_R : [0,\delta_R] \times [0,\delta_{\mu_R}] \rightarrow \mathbb{R}^- \;,
$$
by $P_R(y_0;\mu) = y_1$ where $y_1$ is the first intersection of $\varphi_t^R(0,y_0;\mu)$
with the negative $y$-axis or origin for $t \ge 0$, and
$$ 
	P_L : [-\delta_L,0] \times [0,\delta_{\mu_L}] \rightarrow \mathbb{R}^+
$$
by $P_L(y_1;\mu) = y_2$ where $y_2$ is the first intersection of
$\varphi_t^L(0,y_1;\mu)$ with the positive $y$-axis or origin for $t \ge 0$.

\subsubsection*{Step 2: Show $P_R$ and $P_L$ are well-defined when $\mu=0$}

When $\mu = 0$ the origin is a hyperbolic equilibrium of both \Eq{eq:dhbl} and \Eq{eq:dhbr}; consequently, $P_R(0,0) = P_L(0,0) = 0$. Since when $\mu = 0$ the matrices $A_j$ have eigenvalues \Eq{eq:eigen}, the trajectories of the linearized systems spiral about the origin, repeatedly intersecting the $y$-axis. The spiral behavior is clockwise because in both linear systems $\dot{x}(0,y;0) = y$. The same can be said of trajectories of \Eq{eq:dhbl} and \Eq{eq:dhbr} sufficiently close to the origin because, by Hartman's theorem, there exist neighborhoods about the origin within which \Eq{eq:dhbl} and \Eq{eq:dhbr} are $C^1$ conjugate to their linearizations \cite{Ha60}. Thus when $\mu = 0$, $P_R$ and $P_L$ are well-defined.

\subsubsection*{Step 3:  Show $P_R$ is well-defined for small $\mu > 0$}

When $\mu > 0$, the origin is no longer an equilibrium
of \Eq{eq:dhbr}. The needed properties of $P_R$ can be obtained by considering the local behavior of the trajectory that passes through the origin. This can be deduced by
computing approximations to the flow of \Eq{eq:dhbr}
on the $y$-axis. Since
$$
	\dot{x}(0,y;\mu) = b(\mu)y + o(y) \;,
$$
$b(0)=1$, and $b(\mu)$ is continuous, there is a neighborhood of the origin
such that the positive $y$-axis flows into the right half-plane and
the negative $y$-axis flows into the left half-plane.
Moreover, since
$$
	\dot{y}(0,y;\mu) = q(\mu)\mu + d(\mu)y + o(y)
$$
and since $q(0)=-1$ and $q(\mu)$ is continuous, then for small $\mu>0$, $\dot{y}<0$ for
sufficiently small $y$. Thus the trajectory that passes through the origin, does so
tangent to the $y$-axis and without entering the right half-plane, see \Fig{fig:dhb_pp}.
Since the flow is continuous, there exists a non-empty interval $[0,\delta_R]$ on the positive $y$-axis that maps into the negative $y$-axis. Thus the map, $P_R$, is well-defined for some $\delta_{\mu_R} > 0$. Furthermore $P_R(0;\mu) = 0$ (since our definition of $P_R$
allows for an intersection at $t = 0$). Finally $P_R$ is $C^{k}$ since \Eq{eq:dhbr} is $C^k$.

\InsertFig{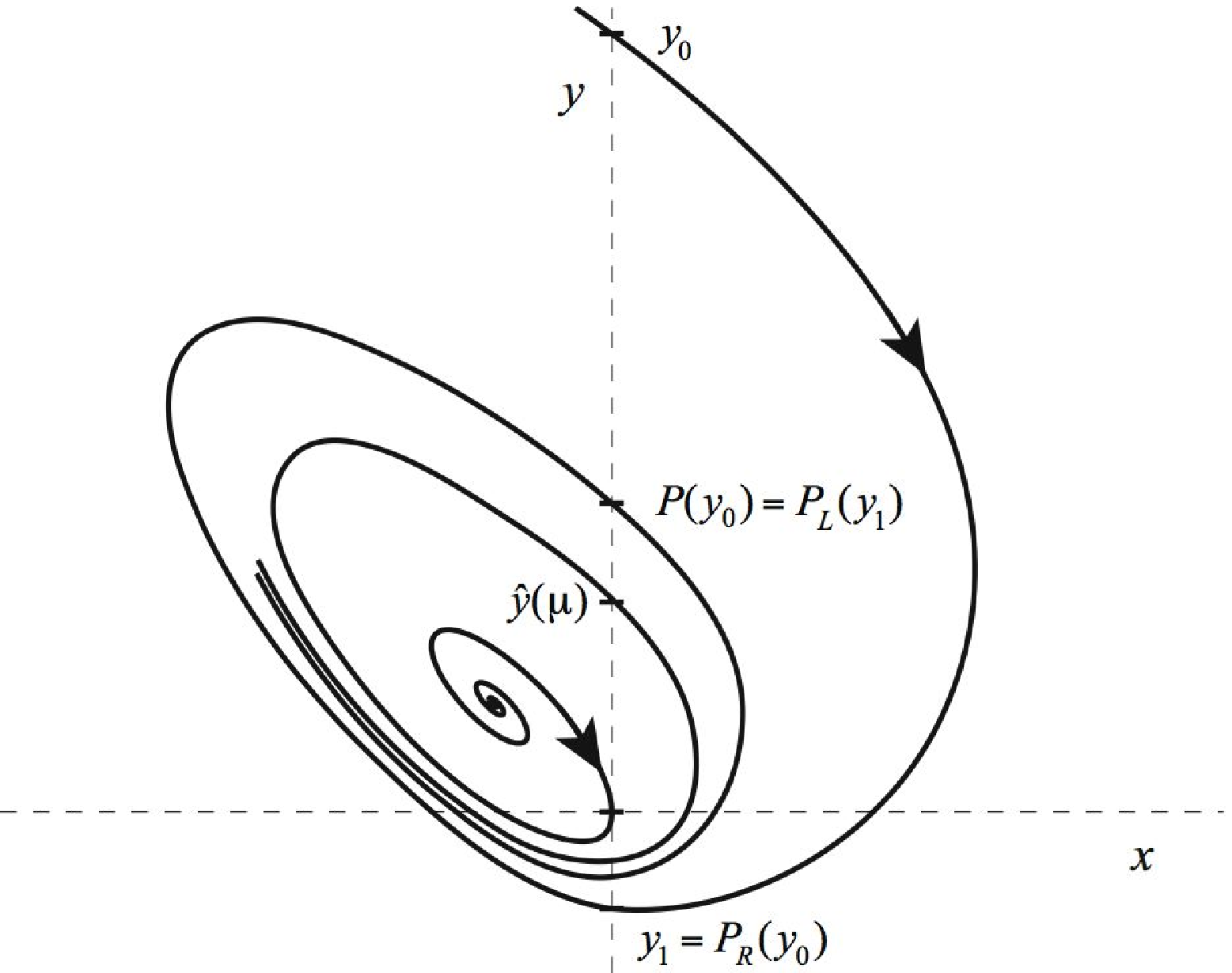}{Sketch of the phase portrait of \Eq{eq:dhb}
when $\mu > 0$}{fig:dhb_pp}{9cm}

\subsubsection*{Step 4: Show $P_L$ is well-defined for small $\mu > 0$}

Since
\begin{equation}
	F_{L}(0,0;\mu) = (0, q(\mu)\mu)
\end{equation}
and $q(0) = -1$, the initial velocity of the
trajectory $\varphi_t^L(0,0;\mu)$ is directly downwards.
The equilibrium of \Eq{eq:dhbl} has the $x$-value given by
\Eq{eq:equil},
and therefore lies in the left half-plane when $\mu$ is small and positive. 

For small $\mu$, the equilibrium lies 
close to the origin so that we may use the linearization of the flow at
the equilibrium to approximate $\varphi_t^L(0,0;\mu)$.
Recall that the eigenvalues of $A_L(0)$ are $\lambda_L \pm i\omega_L$,
where $\lambda_L,\omega_L > 0$, thus $(x_L^*,y_L^*)$
is a repelling focus at $\mu =0$, and by continuity remains
a repelling focus when $\mu$ is small enough. Consequently the flow
$\varphi_t^L(0,0;\mu)$ initially spirals clockwise around the equilibrium
solution within the left half-plane. Since the equilibrium
is repelling, before $\varphi_t^L(0,0;\mu)$ has completed $360^{\circ}$
about the equilibrium it will intersect the $y$-axis at some
point $P(0;\mu) = \hat{y}(\mu) > 0$, see \Fig{fig:dhb_pp}.
Since the flow is continuous, $P_L$ must map a non-empty interval $[-\delta_L,0]$ to points on the positive $y$-axis above $\hat{y}(\mu)$. Thus the map, $P_L$, is well-defined
for some $\delta_{\mu_L} > 0$. Also $P_L$ is $C^{k}$ since \Eq{eq:dhbl} is $C^k$.

\InsertFig{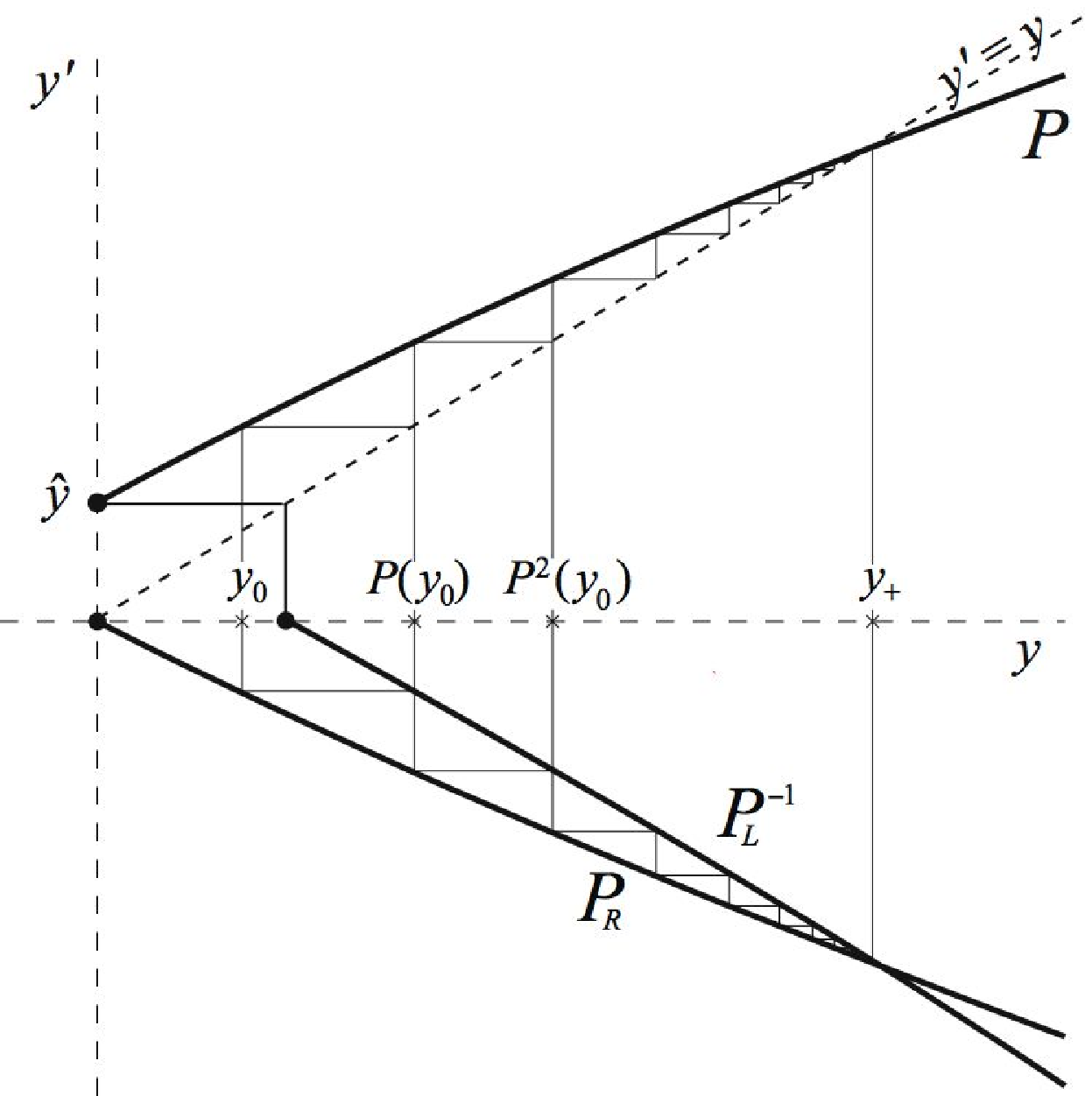}{A sketch of a cobweb diagram for the Poincar\'e map, $P=P_L \circ P_R$, for $\mu > 0$. $P$ has the stable fixed point $y_+$. Also shown are $P_R$ and $P_L^{-1}$.}{fig:web2}{9cm}

\subsubsection*{Step 5: Define $P$ and compute its derivatives at $(0_+,0_+)$}

The results above imply that there are  $\delta, \delta_\mu > 0$ such that the Poincar\'e map, $P:[0,\delta] \times [0,\delta_\mu] \rightarrow \mathbb{R}^+$ defined by 
\begin{equation}\label{eq:poincare}
	P = P_L \circ P_R \;,
\end{equation}
is $C^k$. This map is sketched in \Fig{fig:web2}.

Note that since $P_R(0;\mu) = 0$, its right-sided derivative with respect to
$\mu$ as $\mu \to 0_+$ is $D_\mu P_R(0;0_+) = 0$. Moreover, since 
$P_L(0;\mu) = \hat{y}(\mu)$, and since $\hat{y}(0) = 0$ and and $\hat{y}(\mu) = O(\mu)$
and is positive for $\mu > 0$, we have
$$	
	\alpha \equiv D_\mu P_L(0;0_+) = \lim_{\mu \to 0^+} \frac{\hat{y}(\mu)}{\mu} > 0 \;.
$$
Thus the right-sided derivative of $P$ with respect to $\mu$ is also positive
$$
	D_\mu P(0;0_+) = D_\mu P_L(0;0_+) = \alpha > 0 \;.
$$

To compute the value of $D_y P(0_+;0)$, consider the linear system
$$
	\dot{z} = \left[ \begin{matrix} a & 1 \\ b & d \end{matrix} \right] z \;,
$$
with eigenvalues $-\lambda \pm i \omega$.
For points on the $y$-axis, this system has the flow
$$
	\varphi_t(0,y) = \frac{e^{-\lambda t}}{\omega} 
		\left[ \begin{matrix} \sin(\omega t) \\ 
			-(a+\lambda)\sin(\omega t) + \omega \cos(\omega t) \end{matrix} 
		\right] y  \;.
$$
Consequently it maps the positive $y$-axis onto the negative $y$-axis in time 
$\tau = \frac{\pi}{\omega}$ so that $\varphi_\tau(0,y) = (0,-e^{-\lambda \tau} y)$. 
Moreover, this system approximates the flow of \Eq{eq:dhbr} in the right half-plane 
near the origin. Therefore as 
$y_0 \to 0_+$, $P_R(y_0;0) \to -e^{\frac{-\lambda_R \pi}{\omega_R}} y_0$. 
By a similar argument, as $y_1 \to 0_-$, 
$P_L(y_1;0) \to -e^{\frac{\lambda_L \pi}{\omega_L}} y_1$. Therefore as $y_0 \to 0_+$, 
$$
	P(y_0;0) \to \left( -e^{\frac{\lambda_L \pi}{\omega_L}} \right)
	\left( -e^{\frac{-\lambda_R \pi}{\omega_R}} \right) y_0
	= e^{\Lambda \pi} y_0 \;.
$$
Thus
\begin{equation}\label{eq:DyP}
	D_y P(0_+;0) = e^{\Lambda \pi} \;.
\end{equation}

\subsubsection*{Step 6: Show the periodic orbit exists when $\Lambda <0$}

The function $V:[0,\delta] \times [0,\delta_\mu] \to \mathbb{R}$ defined
by $V(y_0,\mu) = P(y_0;\mu) - y_0$ is $C^k$. Zeros of $V$ correspond to fixed points of the Poincar\'e map, $P$, and thus periodic orbits in the system \Eq{eq:dhb}. In order to apply the implicit function theorem to $V$ at the origin  we must first smoothly extend its definition to a neighborhood of the origin. Let $\tilde{V} : [-\delta,\delta] \times [-\delta_\mu,\delta_\mu] \to \mathbb{R}$ be any $C^{k}$ function with $\tilde{V}(y_0,\mu) = V(y_0,\mu)$ whenever $y_0,\mu \ge 0$.

Note that
\begin{enumerate}
\renewcommand{\labelenumi}{\roman{enumi})}
\item $  \tilde{V}(0,0) = V(0,0) = P(0;0) = 0 $,
\item $ D_y \tilde{V}(0,0) = D_y P(0_+;0) - 1 = e^{\Lambda \pi} - 1 $,
\item $  D_\mu \tilde{V}(0,0)=  D_\mu P(0;0_+) = \alpha > 0$.
\end{enumerate}

Therefore, if $\Lambda \ne 0$, then the implicit function theorem implies that there is a $C^k$ function $y_+$ satisfying $\tilde{V}(y_+(\mu),\mu) = 0$ for all $\mu$ in some neighborhood of $\mu = 0$ such that $y_+(0) = 0$ and 
\begin{equation}\label{eq:yPlusPrime}
   D_\mu y_+(0) = \frac{D_\mu\tilde{V}(0,0)}{ D_y\tilde{V}(0,0)} 
		= \frac{\alpha}{1 - e^{\Lambda \pi}} \;.
\end{equation}

Consequently,  when $\Lambda < 0$, we have $D_\mu y_+(0) > 0$.
Thus for small $\mu > 0$, $y_+(\mu) > 0$ so that $\tilde{V}(y_+(\mu),\mu) =V(y_+(\mu),\mu) = 0$
implying that $y_+(\mu)$ is a periodic point. In other words, the graph of $P(y;\mu)$
necessarily intersects the diagonal as sketched in \Fig{fig:web2}, and when $0 < \mu <\delta_\mu$, \Eq{eq:dhb} has a periodic orbit that intersects the positive $y$-axis at $y_+(\mu)$.

The periodic orbit intersects the negative $y$-axis, at, say, $y_-(\mu) = P_R(y_+(\mu);\mu)$. The function $y_-(\mu)$ is also $C^k$ and vanishes at $\mu = 0$. Moreover, 
\begin{align*}
	D_\mu y_-(0_+) &= D_y P_R(0_+;0) D_\mu y_+(0_+) + D_\mu P_R(0;0_+)\\
				& = \frac{\alpha}{e^{\frac{\lambda_L \pi}{\omega_L}}
					- e^{\frac{\lambda_R \pi}{\omega_R}}} < 0
\end{align*}

\InsertFig{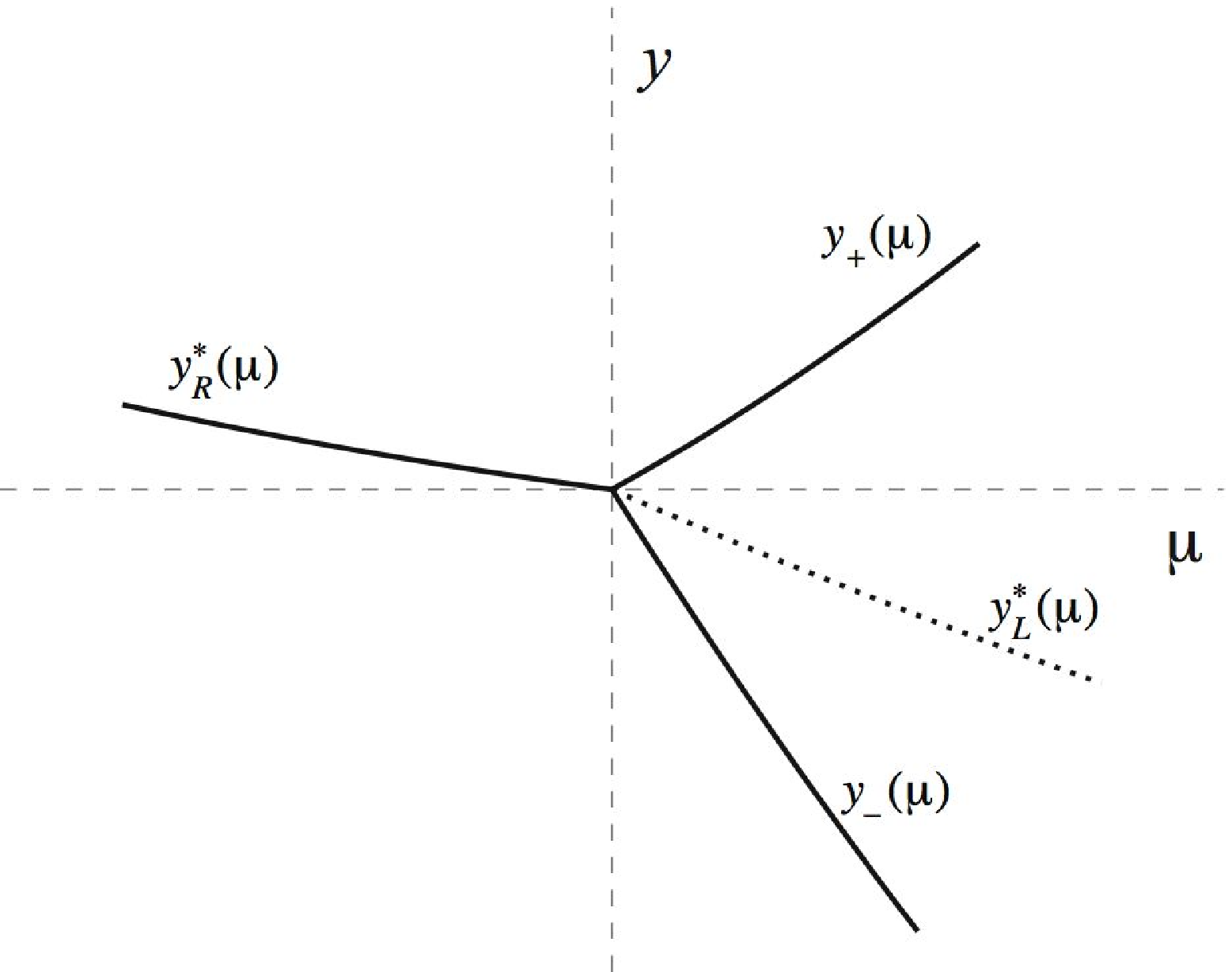}{A sketch of a bifurcation diagram when $\Lambda < 0$.
$y_R^*(\mu)$ and $y_L^*(\mu)$ correspond to the $y$-component of the
equilibrium. $y_+(\mu)$ and $y_-(\mu)$ correspond to the values of the two
intersections of the periodic orbit with the $y$-axis.
Solid [dashed] curves denote attracting [repelling]
solutions.}{fig:dhb_bif}{9cm}

The resulting bifurcation diagram is sketched in \Fig{fig:dhb_bif}. The radius of the periodic orbits, by any sensible definition, grows linearly with respect to $\mu$, to first order.

\subsubsection*{Step 7: Show the periodic orbit is attracting}

The stability of periodic orbits can be deduced by calculating
the value of $D_y P(y_+(\mu);\mu)$. When $\Lambda < 0$, \Eq{eq:DyP} implies that $0 < D_y P(0_+,0)< 1$. Since this function is $C^{k-1}$, we therefore have $0 < D_y P(y_+(\mu);\mu) < 1$, for all sufficiently small $\mu > 0$. Thus the periodic orbit is attracting, as we should expect since the equilibrium is repelling.

\subsubsection*{Step 8: Show there are no periodic orbits for $\Lambda > 0$}

If $\Lambda > 0$, \Eq{eq:yPlusPrime} implies that $y_+^\prime(0) < 0$, so that the locus of zeros, $y_+(\mu)$, fails to enter the positive quadrant ($y_0,\mu > 0$) near $(0,0)$.
Since, by the implicit function theorem, $y_+(\mu)$ is the unique solution that emerges from
the origin, it follows that $\tilde{V}(y_0,\mu) \ne 0$, for all sufficiently small
$y_0,\mu > 0$. Hence in this case, there are no periodic orbits.
\qed

Theorem \ref{thm:dhb} implies that the criticality of the discontinuous Hopf bifurcation depends on $\sgn(\Lambda)$ in \Eq{eq:Lambda}. It is not difficult to understand why this makes sense geometrically. If \Eq{eq:dhb} has a Hopf cycle for small $\mu$, it must encircle the equilibrium and spend time in both the left and right half-planes. Suppose the equilibrium is repelling and lies in the left half-plane. Then, within the left half-plane, the Hopf cycle completes more than $180^\circ$ around the repelling equilibrium of \Eq{eq:dhbl}. However, within the right half-plane it completes less than $180^\circ$ around the inactive, attracting equilibrium of the right vector field \Eq{eq:dhbr}. In order that the orbit be periodic it must, in some sense, spiral outward exactly as much as it spirals inwards. Consequently, the attracting nature of the attracting equilibrium must be stronger than the repelling nature of the repelling equilibrium. Since the time to move $180^\circ$ is $\pi / \omega_j$, this requirement is equivalent to $\lambda_R /\omega_R > \lambda_L / \omega_L$, hence $\Lambda < 0$, in agreement with the theorem.

\InsertFig{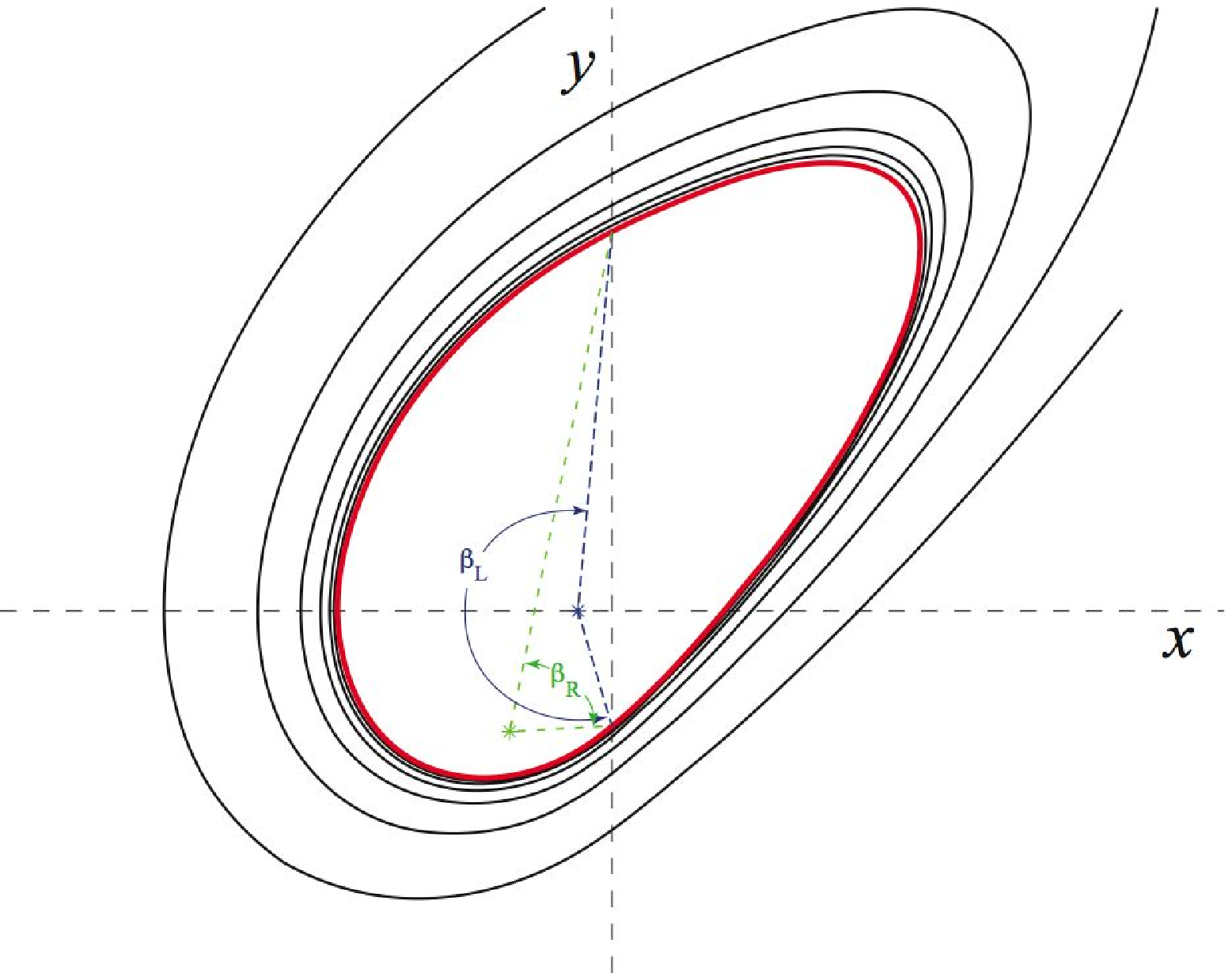}{A trajectory approaching a Hopf cycle for \Eq{eq:example} when $\mu > 0$. The upper equilibrium (blue) is the unstable focus of the left flow and the lower equilibrium (green) is the inactive stable focus of the right flow. The angles that the Hopf cycle subtends within each half-plane about each spiral center are labeled $\beta_L$ and $\beta_R$.}{fig:dhb_orb}{13cm}

In smooth systems, Hopf cycles grow as ellipses and the period
of an arbitrarily small Hopf cycle is easy to calculate \cite{Ku04, Wi03}.
In order to calculate the period for our situation, we must first understand
the shape of the Hopf cycle.
As the Hopf cycle shrinks to a point, it becomes better approximated by the
periodic orbit in the corresponding piecewise-linear system.
A periodic orbit in a piecewise-linear system (like \Eq{eq:example}),
shrinks to zero in a self-similar manner due to an inherent scaling symmetry.
Thus an arbitrarily small Hopf cycle generated by a
discontinuous Hopf bifurcation takes the shape of the periodic orbit in the
corresponding piecewise-linear system, hence the orbit
consists of two spiral segments, see \Fig{fig:dhb_orb}. Again, consider the case $\Lambda <0$ where the equilibrium is repelling and lies in the left half-plane. The Hopf cycle completes, say, $\beta_L > \pi$ radians around the repelling equilibrium and $\beta_R < \pi$ radians around the inactive, attracting equilibrium, see Fig.~\ref{fig:dhb_orb}.
As $\mu \to 0_+$ the ratio $\beta_L /\beta_R$ limits on a 
finite value strictly greater than one.
However, even the knowledge of these angles does not give the period of the cycle since the time taken along each spiral segment is given by two different quantities, $ \gamma_L/ \omega_L$ and $\gamma_R /\omega_R$, respectively. Here we observe that
$0 < \gamma_R < \pi < \gamma_L < 2\pi$, but in general, $\beta_L \ne \gamma_L$ and $\beta_R \ne \gamma_R$. The period of the Hopf cycle is
\begin{equation}\label{eq:period}
	T(\mu) = \left( \frac{\gamma_L}{\omega_L} +
	\frac{\gamma_R}{\omega_R} \right) + O(\mu) \;.
\end{equation}

As a nonlinear example, consider the piecewise continuous system:
\begin{equation}\begin{split}\label{eq:nonlinear}
 \dot{x} &= -x - |x| + y \;,\\
 \dot{y} &= -3x + y - \mu + 3y^2 \;,
\end{split}\end{equation}
which is identical to \Eq{eq:example} except for the addition of a single
nonlinear term. By theorem \ref{thm:dhb}, \Eq{eq:nonlinear} will display the
same local dynamical behavior about the origin for small $\mu$ as \Eq{eq:example}.
\InsertFig{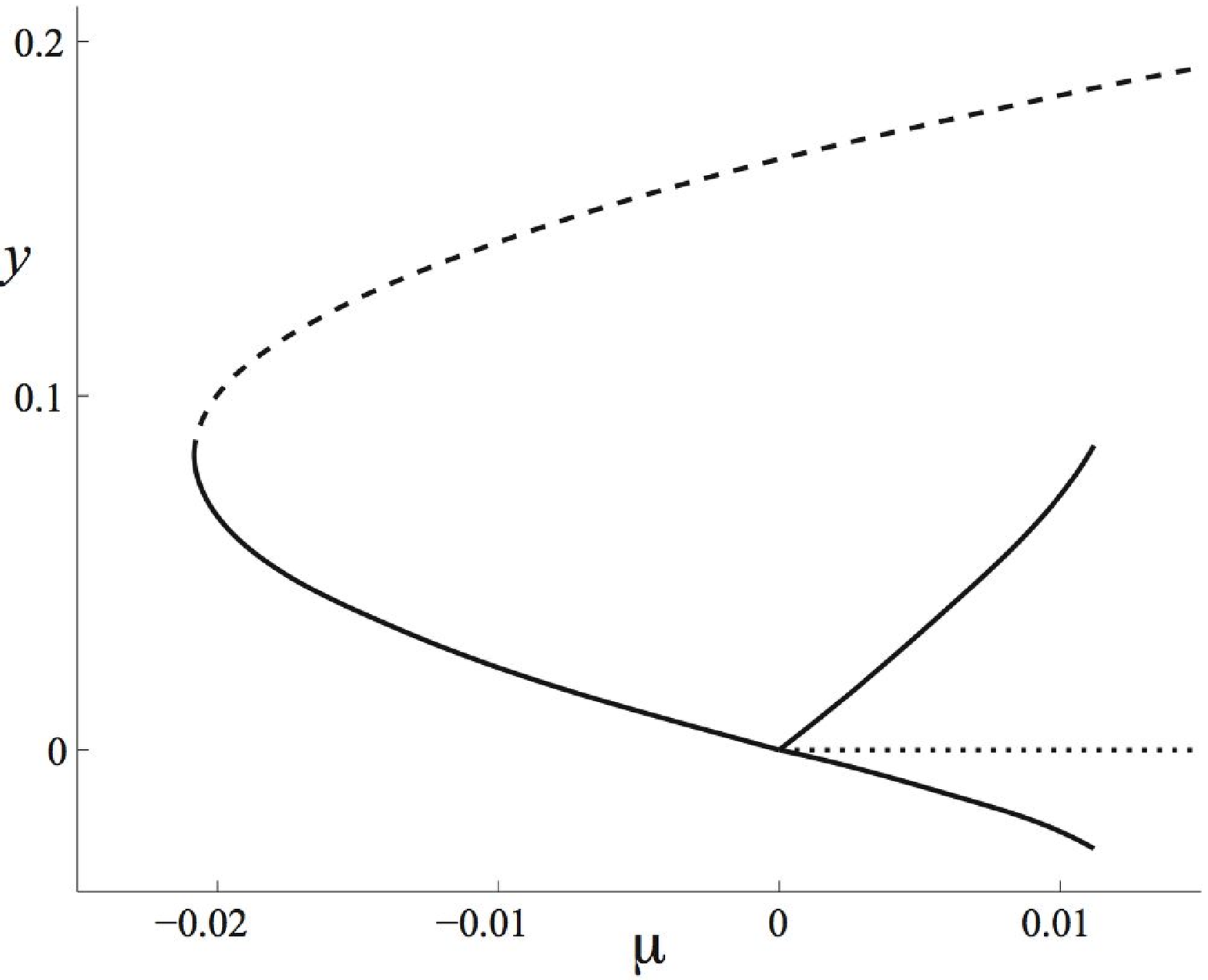}{A bifurcation diagram for \Eq{eq:nonlinear}.
The dotted curve is the unstable focus, and the dashed curve a saddle.
The solid curve, existing for $\mu < 0$, denotes a stable equilibrium.
The solid curves for $\mu > 0$ denote the two locations at which the stable periodic orbit intersects the $y$-axis.}{fig:nonlinearBif}{9cm}
Fig.~\ref{fig:nonlinearBif} shows a bifurcation diagram of \Eq{eq:nonlinear}.
Two equilibria are born in a saddle-node bifurcation at $\mu = -\frac{1}{48} \approx -0.02083$, and exist for all larger values of $\mu$. The stable node becomes a stable focus at $\mu = 2-\frac{7}{6}\sqrt{3} \approx -0.02073$ and finally an unstable focus at $\mu = 0$ when its eigenvalues jump across the imaginary axis from $-\frac12 +
i\omega_L$ to $\frac12 + i\omega_R$, where
$\omega_L = \frac12\sqrt{3}$ and $\omega_R = \frac12\sqrt{11}$. As predicted by \Th{thm:dhb}, a stable periodic orbit is created at $\mu = 0$ exists for small $\mu > 0$. Moreover near the origin, the bifurcation diagram looks the same as that shown in
Fig.~\ref{fig:dhb_bif}.  Phase portraits of \Eq{eq:nonlinear} for $\mu = \pm 0.01$ are shown in Fig.~\ref{fig:nonlinear}.

The periodic orbit is destroyed in a collision with the saddle equilibrium in a homoclinic bifurcation at $\mu \approx 0.01127$.
\InsertFig{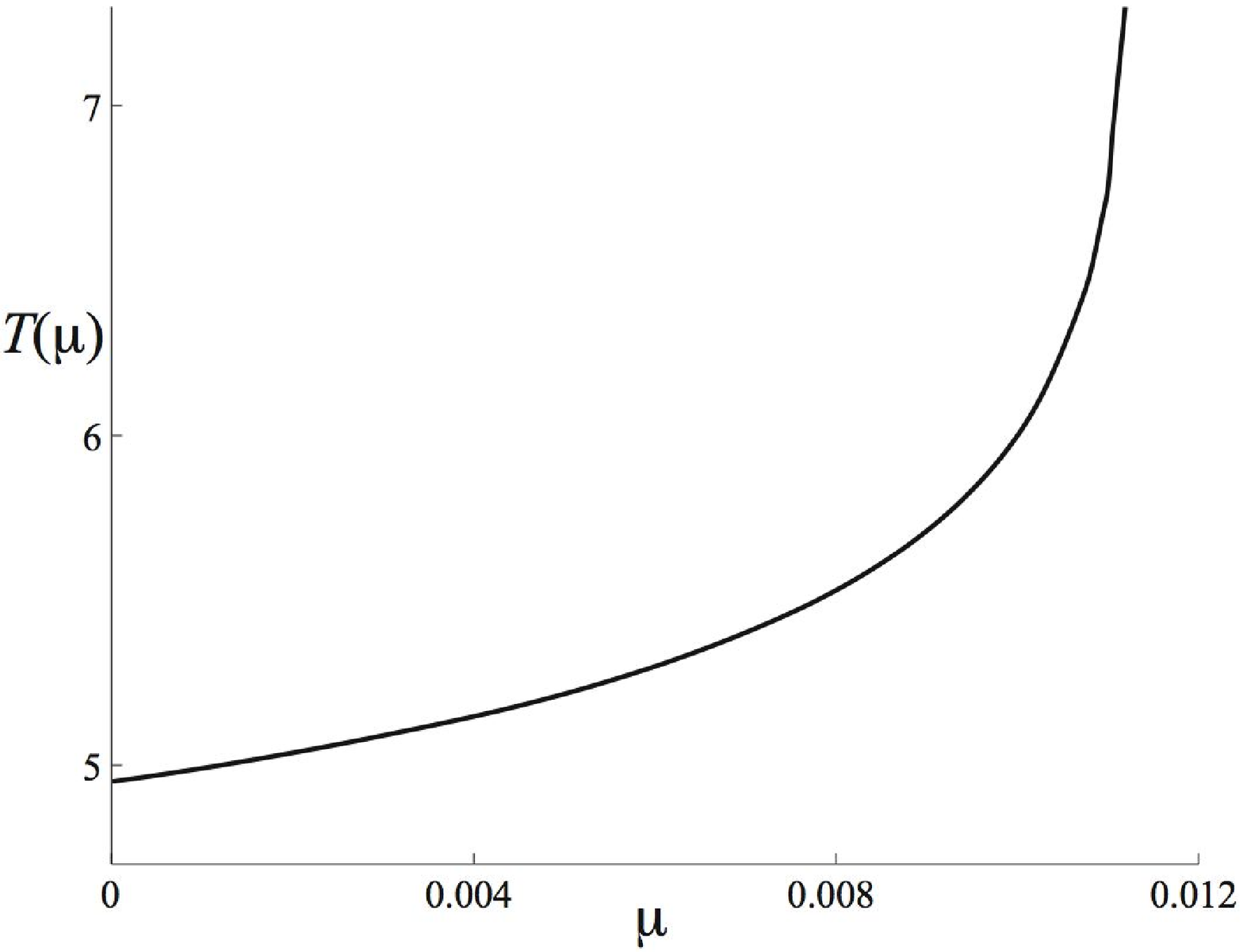}{Period of the Hopf cycle
of \Eq{eq:nonlinear} that is created at the origin when $\mu =0$.}{fig:nonlinearPeriod}{9cm}
The variation of the period with respect to $\mu$ is shown in Fig.~\ref{fig:nonlinearPeriod}. Note that the period at $\mu = 0$ is different from the nominal value, $T(0) \approx 5.522$, that it would have if $\gamma_L = \gamma_R = \pi$ in \Eq{eq:period}. 

\InsertFigTwo{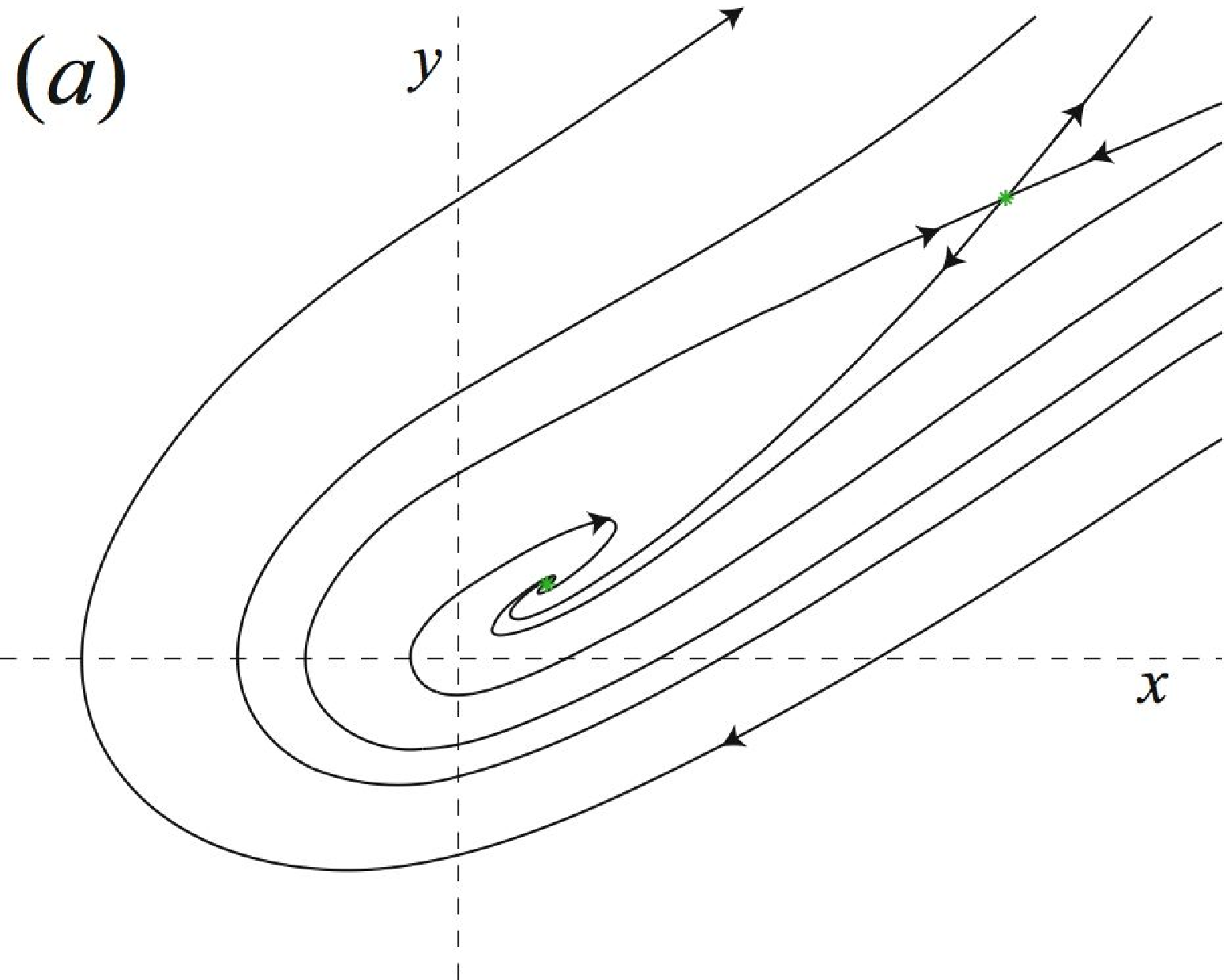}{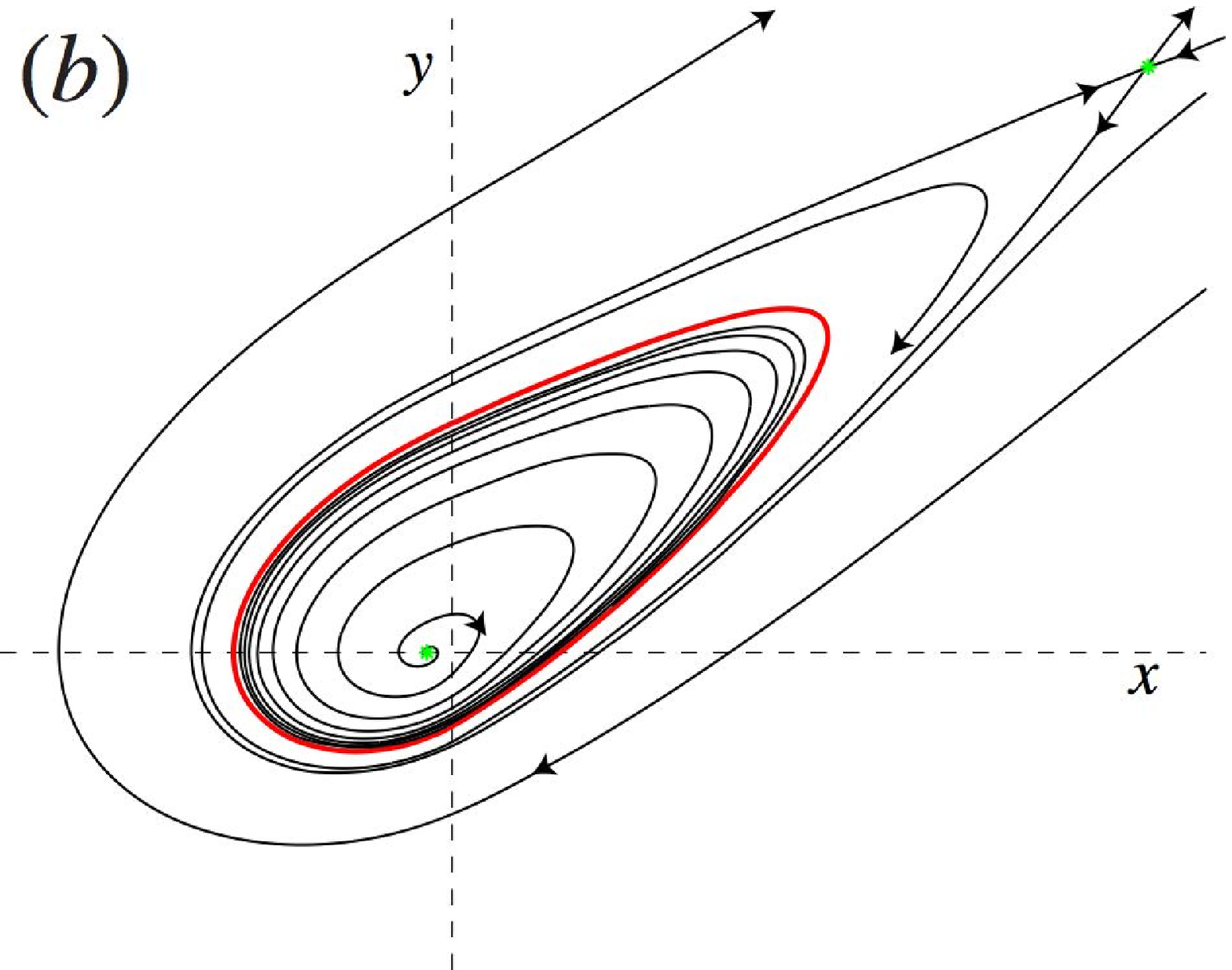}{Phase portraits of \Eq{eq:nonlinear} when
(a) $\mu=-0.01$ and (b) $\mu=0.01$.}{fig:nonlinear}{6cm}

\clearpage

Codimension-two bifurcations can arise if the genericity conditions $\Lambda \neq 0$ and $\Gamma \neq 0$ are not satisfied. As $\Lambda$ crosses zero the bifurcation changes from supercritical to subcritical, and at $\Gamma=0$ the equilibrium does not cross the switching manifold transversely. In both cases, higher order terms are needed for a local analysis. For the special case that the equilibrium is fixed on the switching manifold, $q(\mu) \equiv 0$, a periodic orbit is created if a parameter causes $\Lambda$ to cross zero \cite{ZoKu06}. A further condition required for \Th{thm:dhb} is that the equilibrium intersects the switching manifold at a point where it is smooth. The scenario where this is not true could perhaps be understand by transforming the system so that the switching manifold lies on the positive halves of the $x$ and $y$ axes; a special case where the equilibrium remains at a corner point was treated in \cite{ZoTa05}. For this case, the parameter $\Lambda$ is replaced by a sum of the ratios $\Re(\lambda_j) / \Im(\lambda_j)$ of the eigenvalues multiplied by the opening angle of each sector.

\section{Conclusions and Open Problems}

We have shown there exist Andronov-Hopf-like 
bifurcations in piecewise-smooth continuous systems.
The three major differences between this bifurcation
and the classical Hopf bifurcation are:
\begin{enumerate}
	\item An arbitrarily small Hopf cycle consists of two spiral segments
	as opposed to being elliptical.
	\item The amplitude of the Hopf cycle grows linearly with respect to
	the system parameter instead of as the square root of the parameter value.
	\item The criticality of the bifurcation (i.e., the stability of the Hopf cycle) is determined by linear terms---the parameter $\Lambda$ of \Eq{eq:Lambda}---instead of cubic terms.
\end{enumerate}

It would be nice to extend our results to the case of an $n$-dimensional system with a smooth codimension-one switching manifold such that the spectra of the matrices $A_L$ and $A_R$ differ by one pair of eigenvalues. The difficulty here is devising a version of the center manifold reduction that is used in the proof of the smooth Hopf-bifurcation theorem \cite{MaMc76, Ku04}. The higher dimensional case is complicated by the fact that an equilibrium on a switching manifold can be unstable even when both Jacobian matrices have all of their eigenvalues in the left-half-plane; this was demonstrated for a three dimensional, piecewise linear example by \cite{CaFr06}.

\bibliographystyle{alpha}

\end{document}